\documentclass[aps,preprint,showpacs]{revtex4}%
\usepackage{graphicx}
\usepackage{amsmath}
\usepackage{amsfonts} 
\usepackage{amssymb}%
\begin{document}
\title{Magneto-optics in pure and defective  
Ga$_{1-x}$Mn$_{x}$As from first--principles }
\author{S. Picozzi,$^1$ A. Continenza$^1$, M. Kim$^2$$^,$$^3$  and A. J.
Freeman$^2$} 
\affiliation
{$^1$INFM-CNR, CASTI Regional
Lab. and Dip.  Fisica,
Universit\`a degli Studi dell'Aquila, 67100 Coppito (L'Aquila),
Italy\\
$^2$Department of Physics and Astronomy,
Northwestern University, Evanston,IL 60208
(U.S.A.)\\
$^3$Department of Physics, Seoul National University, Korea}
\begin{abstract}
The  magneto-optical
properties of Ga$_{1-x}$Mn$_{x}$As including their most common defects were 
investigated with precise first--principles density-functional FLAPW
calculations in order to: {\em i})
elucidate the origin of the features in the Kerr 
spectra in terms of the underlying
electronic structure; {\em ii}) perform an accurate comparison with experiments;
and {\em iii}) understand the role of the Mn concentration and occupied sites
 in shaping the
spectra. In the substitutional case, our results show that 
 most of the features have an interband origin and
are only slightly affected by Drude--like contributions, even at low photon
energies. While not strongly affected 
by the Mn concentration for the
intermediately diluted range ($x\sim$ 10\%), 
the Kerr factor shows
a marked minimum (up to 1.5$^o$) occurring at a photon energy of $\sim$
0.5 eV. For  interstitial Mn, the calculated results  bear a striking
resemblance to the experimental spectra, pointing to the comparison
between simulated and experimental
 Kerr angles
as a valid tool to distinguish different defects in the diluted magnetic 
semiconductors framework.
\end{abstract}
\maketitle

The magneto-optical (MO) Kerr effect is known since 1877 when 
Kerr\cite{jkerr}
observed that linearly polarized light reflected from a magnetic
 surface becomes elliptically polarized
 with the major axis  rotated  with respect to 
the incident light. The Kerr effect, as well as  other closely related 
spectroscopic effects,
such as the Faraday effect and magnetic circular dichroism, can be traced back
  to the different interaction of left- and right-
circularly polarized light with a magnetized solid. MO effects are therefore
powerful experimental probes that play a relevant role in clarifying 
the electronic structure of  ferromagnets
and  providing detailed information on the influence of
broken time-reversal symmetry in itinerant quasi-particle electron 
states\cite{review}. 

Within the fascinating field of semiconductor spintronics\cite{science}, transition metal doped semiconductors, due to their 
 unusual
 magnetic and electronic properties, have been the focus of intense research in the last few years.
The early prototype material is undoubtefully Ga$_{1-x}$Mn$_x$As\cite{gamnas}, 
a dilute magnetic semiconductor
(DMS) with $x<$ 10\% and a Curie temperature 
$<$ 150 K, for which its ferromagnetism is generally explained in terms
of a hole-carrier-mediated exchange mechanism.  
Although the electronic and magnetic properties of  Ga$_{1-x}$Mn$_{x}$As have been extensively studied, 
to our knowledge \emph{ab--initio}
 calculations for the MO properties of
Ga$_{1-x}$Mn$_{x}$As have not  been reported. The focus of this work is
therefore to perform an accurate comparison with extensively
performed experiments\cite{munekata,kojima,furdyna}, in order to: \emph{(i)} assess
the validity of the first principles electronic structure theory and \emph{(ii)} to explain the main features in
the MO spectra of GaMnAs on the basis of its electronic structure.

In our  approach based on 
density functional theory (DFT), the Kohn-Sham equations are solved
self-consistently
 using the highly accurate full--potential linearized augmented
 plane wave (FLAPW) \cite{FLAPW} method. Spin-orbit coupling (SOC),
essential to describe
 MO
 effects, neglected in the self--consistent cycle,
  is included as a second
 variational step \cite{soc} in the evaluation of the optical conductivity
 tensor, $\sigma_{ij}$.\cite{mykim}
 For metals (or half-metals, such as GaMnAs), the components of the optical
 conductivity tensor are given by a sum of interband and intraband
 contributions. The interband transitions
  are obtained according to the Kubo formalism within linear
 response theory through momentum matrix 
 elements $\Pi_{ij}$:\cite{ebert},
\begin{eqnarray}
\sigma_{\alpha \beta}(\omega)& =& \frac{i}{\omega V} \int \frac{d{\bf
k}}{(2\pi)^3}
\sum_{i,j} 
(\frac{\Pi^{\alpha}_{ji} \Pi^{\beta}_{ij}}{\omega + i\tau - \epsilon_{ij}} -
\frac{\Pi^{\alpha}_{ij} \Pi^{\beta}_{ji}}{\omega + i\tau + \epsilon_{ij}}),
\hspace{1.5 cm}
\alpha,\beta = x,y,z
\label{MOK.EQ1}
\end{eqnarray} 
where $\omega$ is the photon energy, $\epsilon_{ij}$ are differences
between eigenvalues and $\tau$ is the interband relaxation time.  
The intraband contribution is added to the
diagonal components of the conductivity tensor  with
a phenomenological Drude expression:
$\sigma=\frac{\omega_p^2}{4\pi(1-i\omega\tau_D)}$
where the plasma frequency is given by:
$\omega_p=\frac{4\pi e^2}{m^2 V} \sum_{i{\bf k}} \delta(\epsilon_{i{\bf
k}}-E_F)|\Pi^{\alpha}_{ii}|^2$
($E_F$ is the Fermi energy and $\tau_D$ is the intraband
relaxation time).

Here, we only consider the Kerr effect in the
so-called {\em polar geometry}, in which the incident wave vector and
magnetization are perpendicular to the surface. In this case,
the Kerr rotation angle
 $\theta_{k}(\omega)$ and its ellipticity $\varepsilon_{k}(\omega)$ can be obtained
 from the conductivity tensor as follows\cite{review}:
 \begin{eqnarray}\label{kerreq}
  \theta_{k}(\omega)+i \varepsilon_{k}(\omega)&=&
 -\frac{\sigma_{xy}(\omega)}{\sigma_{xx}(\omega)\sqrt{1+i(4\pi/\omega)\sigma_{xx}(\omega)}}
 \end{eqnarray}
 In order to investigate the effect of Mn concentration on the MO properties, we
 considered different unit cells (each containing a single Mn impurity and with
 the GaAs experimental lattice constant, $a$ = 10.69 a.u.) -- a bcc- and
  an fcc-crystal for a 6.25 \% and a 12.5 \% Mn concentration, respectively.
 We used a basis set of plane waves with wave vector up to
 $K_{max}$=3.5 a.u. and an angular momentum expansion up to
 $l_{max}=8$ for both the potential and charge density. The
 muffin-tin radius, $R_{MT}$, for Mn, Ga and As was chosen equal to 2.1,
 2.3, and 2.3  a.u.,
 respectively. The Brillouin zone (BZ) was sampled using a
 (4,4,4) Monkhorst-Pack\cite{MP} cubic shell, whereas the  optical conductivity was
 computed using up to 216 special {\bf k}-points.
In order to investigate the effects of the most common defects on the
MO properties, we also considered 
32-atom cells with {\em i}) an interstitial Mn,  
{\em ii}) an As-antisite,  {\em i.e.} a substitutional Mn located at the 
origin along with an As-antisite located at $(a/2,a/2,0)$, and {\em iii}) an
interstitial-substitutional dimer coupled antiferromagnetically. Although, 
according to DFT predictions, the 
most energetically
favorable site is the substitutional one,
the possibility for some Mn to occupy the interstitial sites and that
As antisites are formed during severely--out--of--equilibrium growth
cannot be ruled out, as several experiments appear to show.\cite{gallagher}


We first focus on the most diluted systems ($x$ = 6.25\%)  with 
Mn in the substitutional position; these can presumably be well
compared with available experiments\cite{kojima,furdyna}.
Figure \ref{kerrang} (a) shows the real part of the complex Kerr angle with
different values of the broadening (bold
and dashed
lines). The remarkable thing is that, for a small broadening of 
$\hbar/\tau$ = 0.1 eV, the Kerr rotation can reach values of more
than 1.5$^o$: this unexpectedly high Kerr rotation
may open the way to magneto-optical applications using
DMS. However, it is also
evident that the maximum value of the Kerr rotation is markedly
dependent on the smearing value used: already with a larger broadening
($\hbar/\tau$ = 0.3 eV), the Kerr rotation shows a value  
of the order of 0.5
degrees, typically observed in 3$d$ ferromagnets\cite{review}. 
Incidentally, we also note that our calculated spectrum  using 
the smaller broadening is in remarkable
agreement with model calculations\cite{hanki} based on a Kohn-Luttinger Hamiltonian and a 
kinetic exchange interaction -- not only the spectral shape, but also the 
energy position
of the peaks is in good coincidence.

The peculiar shape of
the Kerr rotation, with the strong resonance at $\sim$ 0.5 eV and several other
features at higher energies, can be  fully
explained in terms of the numerator ($\omega \sigma_{xy}$) and
denominator ($\omega D = \omega
\sigma_{xx}(\omega)\sqrt{1+i(4\pi/\omega)\sigma_{xx}(\omega)}$
in Eq.\ref{kerreq}, therefore separating
the  MO and optical contributions,
respectively (see Fig.\ref{kerrang} (d)). 
The deep resonance at  $\sim$ 0.5 eV can be ascribed to the minimum of the
denominator, whereas the other peaks follow the numerator trend, in turn
due to the interplay of the SOC and exchange splitting. The strong resonance at low
energies has therefore an ``optical" origin, whereas the high energy part is
instead due to MO effects. In order to investigate whether the presence of the
strong resonance - given its relatively low energy -
is due to an intraband or interband contribution, the Kerr
rotation without the Drude contribution is also shown (see bold vs thin 
line) in Fig.\ref{kerrang} (b). As expected, the exact position of the peak is
 affected by the Drude contribution (or, equivalently, by the plasma
frequency) but the deep resonance is kept and is therefore  
seen to have mainly an interband origin. 

In order to further investigate the MO properties and, in particular,
the effect of the Mn concentration, 
 we show in 
Fig.\ref{expt} (c) the calculated
Kerr rotation and ellipticity, for the 6.25
\% and 12.5 \% case. 
The overall trend as a function of frequency is quite 
similar for the two concentrations, confirming that in this 
intermediate diluted
regime the concentration of magnetic impurities does not strongly affect the
electronic structure -- consistent with previous reports.\cite{zhao}
In particular, the principal features for
relevant energies $<$ 2 eV are present in both  concentration cases -- the
peaks only differ in their energy positions.
This rather weak dependence on concentration was 
already noted from the experiment by Lang {\em et al.}
\cite{furdyna}  that measured a Kerr rotation that is comparable in size 
upon more-than-doubling
the concentration ({\em i.e.} Kerr spectra were measured
for $x$ = 0.014 and 0.03), 
although some of the features
as a function of energy varied significatively with the concentration.

Moreover, it is of interest to compare our results for the substitutional
case  (see first column in 
Fig.\ref{expt}) with available experimental
spectra, 
measured in the very diluted limit 
($x\sim$ 1-3\%)\cite{furdyna} as
well as for a concentration similar to this work 
($x\sim$ 6\%)\cite{kojima}. Although the 
order of magnitude is consistent, 
there are some discrepancies between theory and
experiment. In particular, the experimental Kerr rotation is
always negative, whereas theoretical spectra show some crossing with the zero
$y$ axis. However, the energy
position of the minimum at $\sim$ 1.5-2 eV and of
the maximum at $\sim$ 2.5-3 eV is  reproduced
in our spectra. 

As for the Kerr ellipticity, there are some similarities
between theory and experiment
regarding the crossing of the zero axis as well as the order of magnitude, but
also some severe discrepancies (especially in the low energy range). 
There might be several reasons for this disagreement, among which we
mention: {\em i}) the poor
description of features in the DFT electronic structure
related to excited or correlated states\cite{notaldau}
 and {\em ii}) the fact that the 
calculations are done at zero
temperature. Indeed, most of the measurements have been done
at higher temperatures, where the magnetization is expected to 
be smaller, therefore
giving rise to a smaller Kerr angle. As a confirmation of that, we recall
that experiments
 by Kojima {\em et al.}\cite{kojima} were performed at 77 K with a Curie 
temperature of 110 K (and, therefore, the measurements were done
at a temperature not much lower than $T_C$); on the other hand,
  experiments by Lang  {\em et al.} 
were performed at 1.8 K with a Curie
 temperature of 60 K and the samples clearly show a  saturated magnetization.
As a result, the Kerr signal in the latter case is higher than in the former case; moreover,
consistent with this temperature dependence,
our results - which ideally reproduce the situation at 0 K or, at least, in the condition of saturated magnetization - predict a rather high value of the Kerr angle.
 
The disorder that might be present in the samples is another reason that could
explain the discrepancy between theory and experiment. Therefore,
in order to investigate the effects of the most common
defects on the MO properties, we also
show in Fig. \ref{expt} the calculated Kerr angles for (b)
the interstitial Mn and (c) substitutional Mn along with an
As-antisite. 
Clearly, different Mn positions or the presence of  
antisites 
significatively change the spectra, pointing to a possible use of the MO Kerr
effect to distinguish  substitutional and interstitial Mn, antisites, etc.
In particular, 
it is immediately
evident that there is a very good agreement as far as the Kerr rotation and
ellipticity are concerned for the interstitial case --- as is evident both with small and large broadenings 
(see solid and dashed line in Fig.\ref{expt}) used to better
reproduce the experiments by Lang\cite{furdyna} and Kojima\cite{kojima}, respectively.
This might be explained as
follows:
 most of the samples\cite{kojima} 
were annealed and it is well known that
annealing brings interstitials to the surface. Therefore, the Kerr technique,
which mostly probes the surface samples, sees an enhanced  contribution from
the interstitials.
As far as the ellipticity only is concerned, we note that a generally 
satisfying agreement is reached also for the As-antisite.
We do not provide plots for the interstitial-substitutional Mn dimers, 
since the agreement is rather poor, as expected. In fact, this is 
compatible with the idea that
the samples in Ref.\cite{kojima} were annealed at 280$^o$ C, and so
the dimers - possibly formed during the low--temperature growth -
are expected to be split and to no longer exist in such an 
appreciable density as to give a strong contribution to the Kerr spectra.

Finally, we point out that our DFT-FLAPW simulations
for hexagonal MnAs, which are in good agreement with previously published
spectra\cite{delin} and
therefore not reported here, show a much larger (on the order of
0.5 degrees over the whole energy range) Kerr rotation  and ellipticity ---
and
strikingly different from experiments for Ga$_{1-x}$Mn$_{x}$As.
Therefore, the comparison between theory and experiment confirms that the Kerr
signal is not due to precipitates in the samples and suggests that this kind of
simulations is a valid means  to reveal competing phases that can frequently
occur during DMS growth\cite{coreani}.


In summary, accurate FLAPW calculations within density functional theory and the
Kubo formalism were performed focusing on the optical and magneto--optical
properties of GaMnAs.
The maximum
Kerr angle,
occurring at a photon energy of $<$0.5 eV,
 can reach high values ($>$ 1.5 $^o$)
and is
mainly due to an optical rather than a magneto-optical origin. 
In the intermediate diluted regime ($x\sim$ 10\%), the Kerr spectra
do not depend dramatically on concentration and are mostly due to interband,
 rather than intraband, contributions.
For Mn in the substitutional position, the comparison
with experiments for the Kerr rotation and ellipticity shows
some disagreement. On the other hand, the spectra are quite well reproduced
for  interstitial Mn, suggesting that the Kerr effect might be used to 
distinguish the MO response of substitutional rather than interstitial Mn.
However, the role that the
somewhat inaccurate treatment of
 correlation or many--body effects within a single-particle DFT description of
 the GaMnAs electronic structure might have in shaping the
Kerr spectra has still to be investigated.

Work at Northwestern University supported by the National Science Foundation 
Grant No. DMR-0244711/002.

\begin{figure}
\caption{(a) Calculated Kerr rotation for Mn in the substitutional site
 for different smearings --- of 0.1 eV (dashed
line) and
0.3 eV (bold solid line) --- as a function of energy. 
Circles denote values from model 
calculations.\protect\cite{{hanki}} (b) Comparison between total (interband +
intraband) contributions (bold solid line) and interband only (thin solid).
(c) Effect of Mn concentration: $x$ = 6.25$\%$ (bold solid line)
and $x$ = 12.5$\%$ (dot-dashed line). (d) Kerr rotation broken down in numerator
({\em i.e.} MO part, dot-dashed line, right $y$-axis) and denominator
({\em i.e.} optical part, solid line, left $y$-axis), see text for details.} 
\label{kerrang}
\end{figure}

\begin{figure}
\caption{Calculated complex Kerr angle (thin solid line)
for an energy broadening of 0.3 eV:  
rotation (upper panels)
and   ellipticity  (lower panels) for (a) 
substitutional Mn, (b) interstitial Mn
and (c) substitutional Mn along with an As antisite. Experimental data
are marked by symbols: circles, diamonds and squares
for Refs.\protect{\cite{kojima}}, \protect{\cite{furdyna}}
and \protect{\cite{munekata}}, respectively. In panel (b), we also show the
 Kerr rotation calculated with a small broadening ($\hbar/\tau$ = 0.1 eV, see
 dashed line).}
\label{expt}
\end{figure}
\end{document}